\documentstyle[preprint,aps]{revtex}

\begin{document}

\title{Thermal noise in
superconducting quantum point-contacts}

\author{A. Mart\'{\i}n-Rodero, A. Levy Yeyati, and F.J.
Garc\'{\i}a-Vidal$^{(\star)}$}

\address{
Departamento de F\'\i sica  Te\'orica de la Materia Condensada C-V.\\
Facultad de Ciencias. Universidad Aut\'onoma de Madrid.\\
E-28049 Madrid (Spain). \\
and, \\
$^{(\star)}$
Condensed Matter Theory Group \\
The Blackett Laboratory \\
Imperial College,
London SW7 2BZ (UK).}

\maketitle

\begin{abstract}
We present a theory for the frequency-dependent current
fluctuations in superconducting quantum point-contacts (SQPC) within the
dc transport regime.
This theory is valid for any barrier transparency between the tunnel and
ballistic limits, yielding an analytical expression for the fluctuations
spectrum in the subgap region.
It is shown that the level of noise in a quasi-ballistic SQPC may have a
huge increase in comparison with the case of a normal contact carrying
the same average current.
The effect of this high level of noise on the actual observability of the
current-phase relation for a ballistic point-contact is discussed in
connection with recent experimental measurements.
\end{abstract}


Present technologies make it possible to fabricate
superconducting point-contacts in the nanometer scale. Examples of these
kind of systems are the recently developed atomic size break-junctions
\cite{breakj} and the split-gate superconductor - two-dimensional
electron gas - superconductor junction of Takayanagi et al.
\cite{Takayanagi}. In both cases the electronic transport takes place
through a reduced number of quantum channels, the contact transmission
being a controllable quantity. These features make these systems very
attractive for testing theoretical models of the superconducting
transport beyond tunnel conditions.

Recently, there have been a number of theoretical works devoted to a
detailed analysis of both the dc and ac response of a single channel
point-contact \cite{gphi,Shumeiko,Averin}. In particular, new
illuminating results have been obtained for the ac current in the
hitherto less understood limit of small bias voltages \cite{gphi,Averin}.
However, little attention has been paid to the effect of thermal
fluctuations in the transport properties of this kind of devices. This
is an important issue both due to its intrinsic interest as a
nonequilibrium phenomena and also because they limit the observability
of the measured characteristics. It is clear that thermal fluctuations
have to be taken into account if a direct comparison between theory
and experimental results is to be carried out. Regarding this last
point, some recent experiments \cite{hol} have shown the deviation
of the measured current-phase relation in a mechanically controllable
break junction with respect to the theoretical predicted one
\cite{kulik}. Some authors have recently pointed out the importance of
thermal fluctuations as a source for this deviation \cite{Omelyanchuk}.

The aim of this communication is to present a theory for the thermal
current fluctuations of a SQPC in the dc regime valid for any
contact transmission.
This theory yields an analytical expression for the
zero-frequency noise which, in the limit of low barrier transparencies,
differs strongly from the standard tunnel theory result \cite{Scal}.
In the opposite limit, i.e. for a ballistic contact, we find that the
current fluctuations {\it diverge} when the supercurrent tends to its maximum
value. We claim that this fact explains the difficulties found
for the experimental observation of the predicted current-phase
relationship for a ballistic contact.

In recent works we have introduced a theoretical approach
for the study of the transport properties of superconducting
nanoscale constrictions
\cite{gphi,joslar}. In this approach the system is
described
by a Hamiltonian written in a site representation, from which the
microscopic Bogoliubov de Gennes equations can be derived \cite{joslar}.
Within this model the normal transmission coefficient through the
constriction can be expressed in terms of microscopic parameters,
allowing to establish a complete correspondence with
other approaches based on scattering theory.

For our present purpose of describing an atomic size contact, it will
be sufficient to analyze the following Hamiltonian \cite{gphi}

\begin{equation}
\hat{H} = \hat{H}_{L} + \hat{H}_{R} + \sum_{\sigma}(t e^{i \phi/2}
\hat{c}^{\dagger}_{L \sigma} \hat{c}_{R \sigma} +
t e^{-i \phi/2} \hat{c}^{\dagger}
_{R \sigma} \hat{c}_{L \sigma}) ,
\end{equation}

\noindent
where $\hat{H}_{L}$ and $\hat{H}_{R}$ are the BCS Hamiltonians for
the uncoupled electrodes (defined as $L$ and $R$) ,
$t$ is the hopping parameter which defines the normal transmission
through the single quantum channel connecting
both electrodes; and $\phi$ is the total superconducting phase
difference between the electrodes.
In the present calculations we shall neglect fluctuations in this
superconducting phase difference and concentrate in the contribution to
the current fluctuations arising from thermal excitation of
quasiparticles.

Within this model, the operator associated with the current through the
contact can be written as

\begin{equation}
\hat{I}(\tau) = \frac{i e}{\hbar} \sum_{\sigma} ( t e^{i \phi/2}
\hat{c}^{\dagger}_{L \sigma}(\tau)
\hat{c}_{R \sigma}(\tau) - t e^{-i \phi/2}
\hat{c}^{\dagger}_{R \sigma}(\tau)
\hat{c}_{L \sigma}(\tau) ) ,
\end{equation}

\noindent
where the different creation and annihilation operators appearing in
Eq. (2) are the usual Heisenberg operators at a given time $\tau$.
Then, the spectral density of the current fluctuations is
defined as

\begin{equation}
S(\omega) = \hbar \int d\tau e^{i \omega \tau} \left[ <\delta \hat{I}
(\tau) \delta \hat{I}(0)> + <\delta \hat{I}(0) \delta \hat{I}(\tau)>
\right]  ,
\end{equation}

\noindent
where $\delta \hat{I}(\tau) \equiv \hat{I}(\tau) - <\hat{I}>$.

For the evaluation of the above averages, we perform a decoupling
procedure which is consistent with the BCS mean field theory.
The  spectrum $S(\omega)$
can then be expressed in terms of the single-particle non-equilibrium
Green functions \cite{keld},
$\hat{G}^{+,-}_{\alpha \beta}(\Omega)$ and
$\hat{G}^{-,+}_{\alpha \beta}(\Omega)$ (where
$\alpha$ and $\beta$ can be either $L$ or $R$).
In a superconducting broken symmetry
representation \cite{nambu} $\hat{G}^{+-}_{\alpha \beta}(\Omega)$
is defined by

\begin{equation}
\hat{G}^{+-}_{\alpha \beta}(\Omega)=  \int d\tau e^{i \Omega \tau}
\hat{G}^{+-}_{\alpha \beta}(\tau,0) ,
\end{equation}
\noindent
with

\[ \hat{G}^{+-}_{\alpha \beta}(\tau,0)= i
\left(
\begin{array}{cc}
<\hat{c}^{\dagger}_{\beta \uparrow}
(0) \hat{c}_{\alpha \uparrow}(\tau)>   &
<\hat{c}_{\beta \downarrow}(0)
\hat{c}_{\alpha \uparrow}(\tau)>  \\
<\hat{c}^{\dagger}_{\beta \uparrow}(0)
\hat{c}^{\dagger}_{\alpha \downarrow}(\tau)>  &
<\hat{c}_{\beta \downarrow}(0)
 \hat{c}^{\dagger}_{\alpha \downarrow}(\tau)>
\end{array}  \right) ,  \]

\noindent
and
$\hat{G}^{-+}_{\alpha \beta}(\tau,0)=
[ \hat{G}^{+-}_{\beta \alpha}(0,\tau) ]^{\dagger}$.

In terms of the functions $\hat{G}^{+-}_{\alpha \beta} (\Omega)$
and $\hat{G}^{-+}_{\alpha \beta} (\Omega)$, $S(\omega)$ adopts
the form

\begin{eqnarray}
&&S(\omega)  = \frac{e^2}{\hbar} \int d\Omega
\; \mbox{Tr} [\hat{t} \hat{G}^{+-}_{RL}(\Omega)
\hat{G}^{-+}_{RL}(\Omega+\omega)\hat{t}+ \nonumber \\  &&
\hat{t} \hat{G}^{+-}_{LR}(\Omega)
\hat{G}^{-+}_{LR}(\Omega+\omega)\hat{t}-
\hat{t} \hat{G}^{+-}_{LL}(\Omega)
\hat{G}^{-+}_{RR}(\Omega+\omega)\hat{t}- \nonumber \\  &&
\hat{t} \hat{G}^{+-}_{RR}(\Omega)
\hat{G}^{-+}_{LL}(\Omega+\omega)\hat{t}+
\hat{t} \hat{G}^{+-}_{RL}(\Omega+\omega)
\hat{G}^{-+}_{RL}(\Omega)\hat{t}+ \nonumber \\  &&
\hat{t} \hat{G}^{+-}_{LR}(\Omega+\omega)
\hat{G}^{-+}_{LR}(\Omega)\hat{t}-
\hat{t} \hat{G}^{+-}_{LL}(\Omega+\omega)
\hat{G}^{-+}_{RR}(\Omega)\hat{t}- \nonumber \\  &&
\hat{t} \hat{G}^{+-}_{RR}(\Omega+\omega)
\hat{G}^{-+}_{LL}(\Omega)\hat{t} ] ,
\end{eqnarray}

\noindent
where $ \hat{t} $ is the hopping interaction between
the electrodes written in the ($2 \times 2 $)  Nambu representation

\begin{equation}
\hat{t} =
\left(
\begin{array}{cc}
t e^{i\phi} & 0 \\ 0 & -t e^{-i \phi}
\end{array}  \right) .
\end{equation}

In the present communication we concentrate in the zero
voltage case in which the average current is due to
Cooper pairs. For the calculation of the nonequilibrium Green functions
appearing in Eq. (5) we can the use the relations \cite{joslar}

\begin{eqnarray}
\hat{G}^{+,-}_{\alpha,\beta}(\Omega) = \left[ \hat{G}^{a}_{\alpha \beta}
(\Omega) - \hat{G}^{r}_{\alpha \beta}(\Omega) \right] f(\Omega)
\nonumber \\
\hat{G}^{-,+}_{\alpha,\beta}(\Omega) = -\left[ \hat{G}^{a}_{\alpha
\beta}(\Omega) - \hat{G}^{r}_{\alpha \beta}(\Omega) \right]
[1-f(\Omega)] ,
\end{eqnarray}
\noindent
where $f(\Omega)$ is the Fermi
factor and $ \hat{G}^{r,(a)}_{\alpha \beta}$ are the retarded (advanced)
Green functions of the coupled contact.
These last quantities can be obtained up to infinite order
in the coupling parameter $t$ by solving the following Dyson equation

\begin{equation}
\hat{G}^{r,(a)}_{\alpha \beta}(\Omega) = \hat{g}^{r,(a)}_{\alpha \beta}
(\Omega) \delta_{\alpha \beta} +
\sum_{\gamma} \hat{g}^{r,(a)}_{\alpha \alpha}
(\Omega) \hat{\Sigma}^{r,(a)}_{\alpha \gamma}
\hat{G}^{r,(a)}_{\gamma \beta}(\Omega),
\end{equation}
\noindent
where $\hat{\Sigma}^{r,(a)}_{LL}=\hat{\Sigma}^{r,(a)}_{RR}=0$ and
$\hat{\Sigma}^{r,(a)}_{LR}= \left( \hat{\Sigma}^{r,(a)}_{RL} \right)^*
=\hat{t}$.
The indexes $\alpha$, $\beta$
and $\gamma$ can be either $L$ or $R$, and
$\hat{g}^{r,(a)}_{\alpha \alpha}$ are the retarded (advanced) Green functions
corresponding to the left and right uncoupled electrodes.

For the symmetric case, both electrodes have
the same modulus of the superconducting order parameter, $\Delta$,
and these Green functions can be expressed as

\begin{eqnarray}
\hat{g}^{r,(a)}_{L L}(\omega) = \hat{g}^{r,(a)}_{R R}(\omega) &
= & \frac{1}{W \sqrt{\Delta^2 - (\omega \pm i \eta)^2}} \times\nonumber \\ &&
\left( \begin{array}{cc} -\omega \pm i \eta  & \Delta \\
\Delta & -\omega \pm i \eta \end{array} \right),
\end{eqnarray}
\noindent
where $W$ is an energy scale related to the normal density of states
at the Fermi level by $\rho(\epsilon_F) = 1/(\pi W)$ and $\eta$ is
a small energy relaxation rate that takes into account the damping
of the quasi-particle states due to inelastic processes inside the
electrodes. This parameter can be estimated from the electron-phonon
interaction to be a small fraction of $\Delta$ \cite{Scal2}.
It is useful to define the normal transmission coefficient of the
contact, which in terms of $W$ and $t$ has the form $\alpha = 4
(2t/W)^2/(1+(2t/W)^2)^2$ \cite{contact}. The spectral densities
that are obtained from Eq. (8) are no longer
singular at the gap edges and exhibit poles inside the superconducting
gap, located at energies $\omega_S = \pm \Delta \sqrt{1 - \alpha
\sin^2(\phi/2)}$, corresponding to the interface bound states
\cite{bound}. As stated in previous works, these bound states
carry all the Josephson current in the limit of a short constriction
\cite{joslar,short}. Therefore, their contribution to the
zero-voltage current fluctuations can be expected to
be crucial, as is certainly found.

Once the single particle Green functions are known, the spectrum
$S(\omega)$ can be calculated using Eq. (5).
The typical form of this spectrum is illustrated
in Fig. 1, where $S(\omega)$ is plotted for fixed temperature and
three different contact transmissions.
Notice that for $\omega < 2 \Delta$
the spectrum is formed by two resonant peaks at $\omega
= 0$ and $\omega = 2 \omega_S$, arising from the existence of the
bound states at $\omega_S$. Qualitatively, the peak at zero frequency
increases with increasing transmission, while the one at $2 \omega_S$ is
negligible for both nearly perfect and very small transmissions, adopting
its maximum value around $\alpha \sim 2/3$.
For $\omega > \Delta + |\omega_S|$ contributions
from the continuous part of the single particle spectrum
become important.

In the limit of a very weakly damped contact, i.e. $\eta \ll \alpha
\Delta$,  it is possible to evaluate $S(\omega)$
at $\omega = 0$ and $\omega = 2 \omega_S$  analytically.
We find

\begin{equation}
S(0) = \frac{2 e^2}{h} \frac{\pi}{\eta} \frac{\Delta^4 \alpha^2
\sin^2(\phi)}{\omega_S^2} f(\omega_S) \left[ 1 - f(\omega_S) \right] ,
\end{equation}
\noindent
and

\begin{equation}
S(2 \omega_S) = \frac{2 e^2}{h} \frac{\pi}{\eta} \frac{\Delta^4 \alpha^2
(1 - \alpha)
\sin^4(\frac{\phi}{2})}{\omega_S^2} \left[ f(\omega_S)^2  + f(-\omega_S)^2
\right] .
\end{equation}

These expressions clearly display the important role played by the
interface bound states in fixing the magnitude of the current
fluctuations for subgap frequencies. It should be stressed that, although
the absolute size of the current fluctuations depend on the estimated
value of parameter $\eta$, its precise variation with the superconducting
phase difference and temperature
is controlled only by the contact transmission $\alpha$.

Our analytical results are strictly valid in the limit
$\eta \ll \alpha \Delta$ and differ strongly from the equilibrium
fluctuations obtained using standard tunnel
theory \cite{Scal}, which yields $S(0) \sim \alpha (1 + \cos(\phi))
\ln \Delta/\eta$. This last expression
becomes accurate just in the opposite limit, $\eta \gg \alpha \Delta$
which holds in the tunnel regime, i.e. $\alpha \ll 1$.
In ref. \cite{gphi} we have explicitly shown that the limits $\eta
\rightarrow 0$ and $\alpha \rightarrow 0$ do not commute.
This behavior can be understood in the following way: when $\eta
\ll \alpha \Delta$, multiple Andreev scattering processes
give the dominant contribution
to any dynamical quantity and should be included up to infinite order.
On the other hand,
when $\alpha$ is small enough (in such a way that $\alpha \Delta \ll
\eta$) these high order
scattering events become heavily damped and the lowest term of the perturbative
expansion in $t$ gives the correct result.
For a realistic SQPC in which, as commented above,
$\eta$ can be estimated to be a small fraction of $\Delta$,
the situation would always correspond to the weakly
damped regime, except for extremely small values of $\alpha$ and
therefore Eqs. (10)  and (11) will accurately describe the low frequency noise.

The analysis of Eq. (10) reveals some remarkable physical consequences.
To begin with, and in contrast to the normal case where a reduction
of noise is found \cite{normal}, $S(0)$
experiences a dramatic increase when approaching the ballistic regime.
More precisely, there is a value of $\alpha$, given roughly by
the condition
$k_B T \sim \Delta \sqrt{1 - \alpha}$, above which there is an exponential
increase of the thermal noise. On the other hand, in this last
situation, there appears a very strong asymmetry on the phase-dependence
of $S(0)$, with its maximum value progressively moving from $\phi =
\pi/2$ to $\phi = \pi$.
This remarkable behavior should certainly have implications in the
actual observability of the
supercurrent-phase relation in a SQPC.

In order to analyze the importance of these thermal fluctuations
it is convenient
to study the ratio $S(0)/2 e <I(\phi)>$, where $<I(\phi)>$ is the
phase-dependent average supercurrent (let us recall that for a normal
contact the classical shot noise limit corresponds to $S(0) = 2 e <I>$),
given by \cite{joslar}

\begin{equation}
<\hat{I}(\phi)> = \frac{e \pi}{h}  \frac{\Delta^2 \alpha
\sin(\phi)}{\mid \omega_S \mid} \tanh[\frac{\mid \omega_S \mid}{2k_{B}T}] .
\end{equation}

In Fig. 2 we plot $S(0)/2 e <I(\phi)>$ as a function of the
superconducting phase difference
for increasing values of the transmission $\alpha$ and two
different temperatures. This figure illustrates the huge increase of
thermal noise when the transmission becomes sufficiently large.
As can be observed, for a reasonable choice of parameter $\eta$
and depending on the temperature, the level of noise can reach values
several orders of magnitude larger than $2 e <I(\phi)>$.
When lowering the temperature this level of noise is reduced, but it
will always be significant close to the ballistic case in a phase
interval around $\phi = \pi$,  just in the zone where the average
current has its maximum at low temperatures.
In fact, taking the
limit $\alpha \rightarrow 1$, the ratio $S(0)/2 e <I(\phi)>$
has the form

\begin{equation}
\frac{S(0)}{2e<\hat{I}(\phi)>}(\alpha \rightarrow 1) =
\frac{2 \Delta}{\eta} \frac{\sin[\frac{\phi}{2}]}
{\sinh [ \frac{\Delta \cos[\frac{\phi}{2}]}{k_B T} ]} ,
\end{equation}
\noindent
which clearly diverges when $\phi$ approaches $\pi$.
The fact that the zero frequency noise has large values in the
zone where the maximum of the average current occurs
can explain the experimental difficulties found to
observe the predicted $\sim \sin{\phi/2}$
form of the current-phase relation for junctions with
direct conductivity, as reported in ref. \cite{hol}.

Finally, it is worth discussing our result for the zero frequency
noise of a weakly damped contact in the light of the Callen-Welton
fluctuation-dissipation theorem. In general, this theorem relates
the equilibrium current fluctuations with the
linear conductance $G$ by $S(0)=4 k_{B} T G$.
This relation allows us to
calculate in a straightforward way the phase-dependent
linear conductance of a SQPC from Eq. (10)

\begin{equation}
G(\phi) = \frac{2 e^2}{h} \frac{\pi}{k_{B}T \eta} \left[
\frac{\Delta^2 \alpha \sin(\phi)}{4 \omega_S}
\mbox{sech} (\frac{\omega_S}{2 k_BT}) \right]^2.
\end{equation}

Eq. (14) coincides exactly with the result of ref. \cite{gphi} in
which a direct calculation of the linear conductance of a
SQPC was performed (this expression for
$G(\phi)$ has been recently rederived in \cite{Averin2} for the
particular case of a ballistic contact).

In conclusion, we have developed a theory
of the thermal fluctuations for a SQPC in the dc regime.
The noise spectrum exhibits resonant peaks at subgap frequencies
associated with the existence of bound states in the constriction
region.
For the case of a weakly damped contact
($\eta \ll \alpha \Delta$), we have obtained
a closed analytical expression for the weight of these resonant peaks.
We have shown that a striking consequence of
the presence of these bound states is a huge increase of the low
frequency noise level when approaching the ballistic limit,
this high level of noise being particularly important when the
average supercurrent is close to its maximum value.
We claim that these results may explain
the reported difficulties in measuring the
predicted $\sin{\phi/2}$ behavior for $<I(\phi)>$
in the case of highly transmissive contacts. Finally, we have discussed
the connection between the present theory and the phase-dependent
conductance of a SQPC by means of
the fluctuation-dissipation theorem.

Support by Spanish CICYT (Contract No. PB93-0260) is acknowledged.
One of us (A.L.Y.) acknowledges support by the European
Community under contract No.CI1*CT93-0247.

\begin{figure}
\caption{Current fluctuation spectrum of a SQPC in the dc regime for
three different values of the transmission. The superconducting phase
difference corresponds in each case to the maximum supercurrent and the
temperature is $k_BT = 0.2 \Delta$.}
\end{figure}

\begin{figure}
\caption{The ratio between the zero frequency noise and the average
supercurrent as a function of the superconducting phase difference for
increasing values of the transmission coefficient.}
\end{figure}


\begin{thebibliography}{10}

\bibitem{breakj} N. van der Post, E.T. Peters, I.K. Yanson and
J.M. van Ruitenbeek, Phys. Rev. Lett. {\bf 73}, 2611 (1994);
B.J. Vleeming, C.J. Muller, M.C. Koops, and R. de Bruyn Ouboter,
Phys. Rev. B {\bf 50}, 16741 (1994).

\bibitem{Takayanagi} H. Takayanagi, T. Akazaki and J. Nitta, Phys. Rev.
Lett. {\bf 75}, 3533 (1995).

\bibitem{gphi} A. Mart\'{\i}n-Rodero, A. Levy Yeyati and J.C. Cuevas,
Physica B, in press; A. Levy Yeyati,  A. Mart\'{\i}n-Rodero and J.C.
Cuevas, J. Phys.: Condens. Matter, in press (cond-mat 9505102).

\bibitem{Shumeiko} E.N. Bratus, V.S. Shumeiko and G. Wendin, Phys. Rev.
Lett. {\bf 74}, 2110 (1995).

\bibitem{Averin} D. Averin and A. Bardas, Phys. Rev. Lett. {\bf 75},
1831 (1995).

\bibitem{hol} M.C. Koops, L. Feenstra, B.J. Vleeming,
A.N. Omelyanchouk and R. de Bruyn Outober, to be published in
Physica B (1995).

\bibitem{kulik} O. Kulik and A.N. Omelyanchouk, Fis. Nisk. Temp.
{\bf 3}, 945 (1977);{\bf 4}, 296 (1978) [Sov. J. Low Temp. Phys.
{\bf 3}, 459 (1977);{\bf 4}, 142 (1978)].

\bibitem{Omelyanchuk} R. de Bruyn Ouboter and A.N. Omelyanchouk,
unpublished.

\bibitem{Scal} D. Rogovin and D. J. Scalapino, Annals of Physics {\bf 86},
1 (1974).

\bibitem{joslar} A. Mart\'{\i}n-Rodero, F. J. Garc\'{\i}a-Vidal and
A. Levy Yeyati, Phys. Rev. Lett. {\bf 72}, 554 (1994);
A. Levy Yeyati, A. Mart\'{\i}n-Rodero, F. J. Garc\'{\i}a-Vidal,
Phys. Rev. B {\bf 51}, 3743 (1995).

\bibitem{keld} L. V. Keldysh, Sov. Phys. JETP {\bf 20}, 1018 (1965).

\bibitem{nambu} Y. Nambu, Phys. Rev. {\bf 117}, 648 (1960).

\bibitem{Scal2} A typical estimate of $\eta$ for a traditional
superconductor is $\eta/\Delta \sim 10^{-2}$. See for instance, S.B.
Kaplan et al., Phys. Rev. B {\bf 14}, 4854 (1976).

\bibitem{contact} J. Ferrer, A. Mart\'{\i}n-Rodero and F. Flores,
Phys. Rev. B {\bf 38}, 10113 (1988).

\bibitem{bound} A. Furusaki and M. Tsukada,
Physica B {\bf 165 \& 166}, 967 (1990).

\bibitem{short} C. W. J. Beenakker and H. van Houten, Phys. Rev. Lett.
{\bf 66}, 3056 (1991).

\bibitem{normal} V.A.Khlus, Zh.Eksp.Teor.Fiz. {\bf 93}, 2179 (1987)
[Sov. Phys. JETP {\bf 66}, 1243 (1987)];
G.B.Lesovik, Pis'ma  Zh.Eksp.Teor.Fiz. {\bf 49},
594 (1989) [JETP Lett. {\bf 49}, 683 (1989)];
M.Buttiker, Phys.Rev.Lett. {\bf 65}, 2901 (1990).

\bibitem{Averin2} D. Averin and A. Bardas, cond-mat 9509153.
\end{thebibliography}
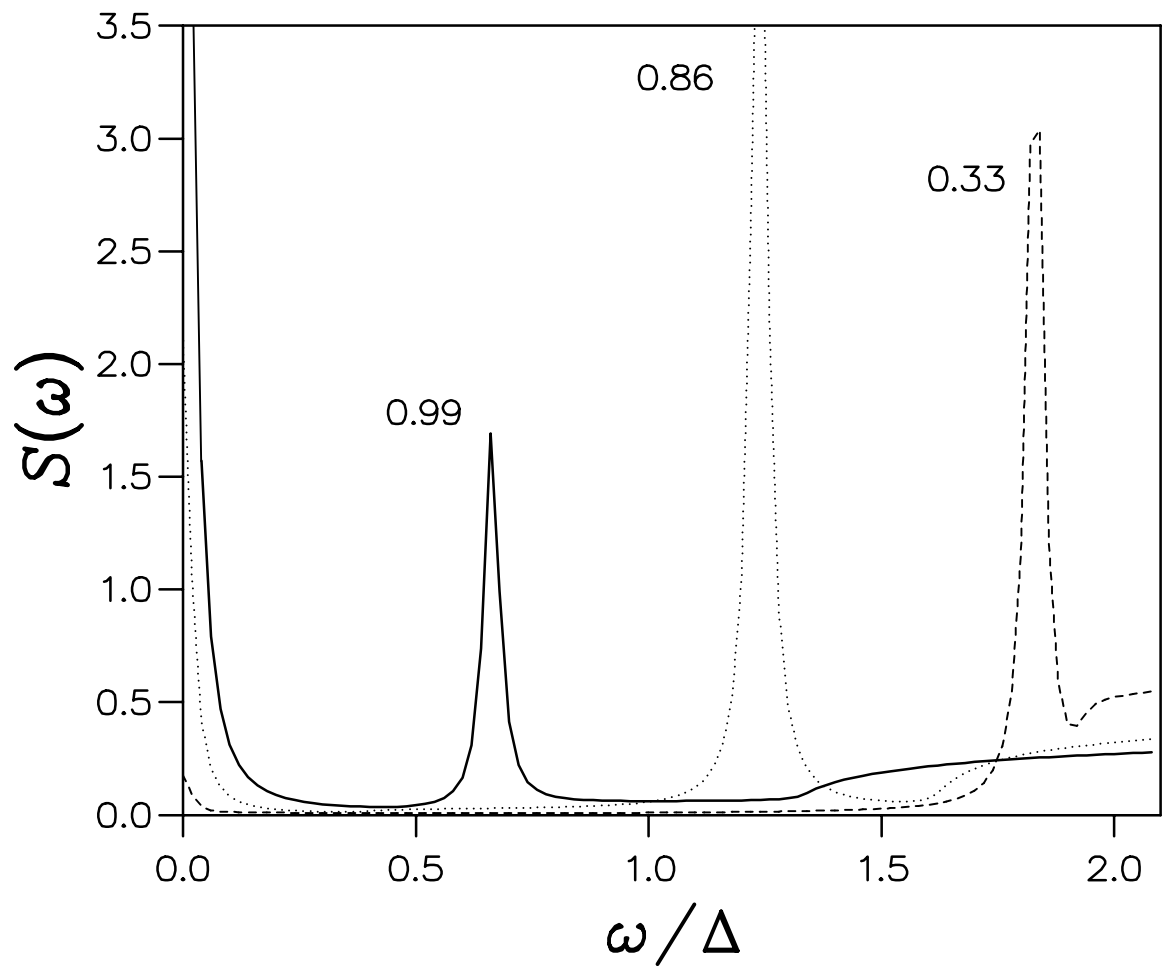
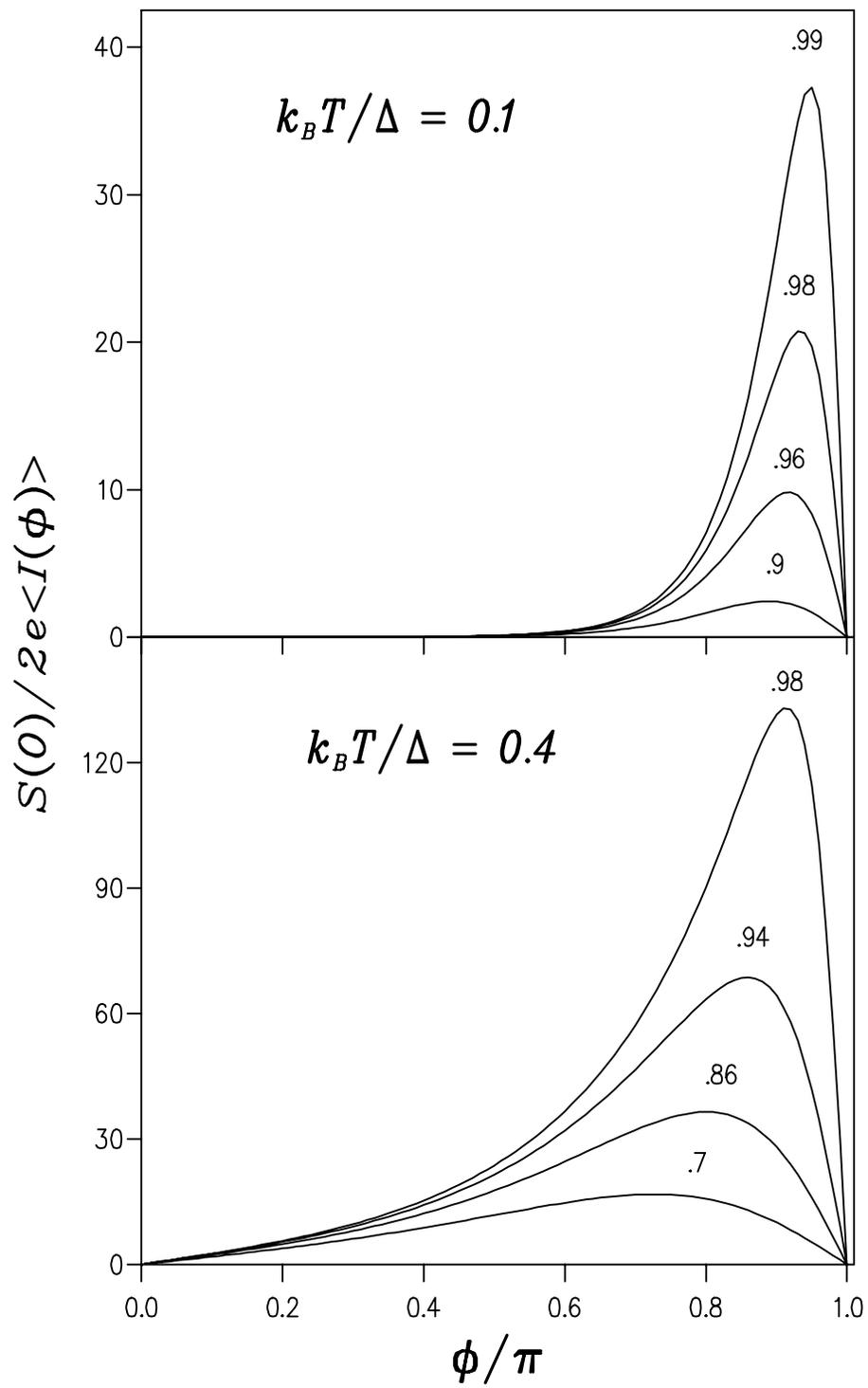
\end{document}